# The new methods for equity fund selection and optimal portfolio construction

Yi Cao

Bank of Jiangsu Asset Management, China

[caoyi1@jsbchina.cn]

*Abstract*
*We relook at the classic equity fund selection and portfolio construction problems from a new perspective and propose an easy-to-implement framework to tackle the problem in practical investment. Rather than the conventional way by constructing a long only portfolio from a big universe of stocks or macro factors, we show how to produce a long-short portfolio from a smaller pool of stocks from mutual fund top holdings and generate impressive results. As these methods are based on statistical evidence, we need closely monitoring the model validity, and prepare repair strategies.*

## I Introduction
The asset management arm of commercial banks in China is growing its presence in equity investment. It faces the challenging problem of constructing fixed income type portfolios providing stable and high return for retail investors, who used to invest solely in fixed income in past many years. The existing players in the market are mutual funds and hedge funds.

However, the classic ways to study portfolio management under the Markowitz's mean variance framework suffers big problems in practical investment, ie: relying on metrics such as mean, variance and covariance matrix ignores the time series structure of the prices which are heterogeneous over time. The modern factor models such as Fama's 3 factor model or Barrar's risk models suffer from problems of non-investable factors and variant factor loaders (beta).

More recent fancy application of machine learning techniques, eg deep learning and reinforced learning in finance is criticized for complication, overfiting and prone to breakdown. Many are impractical and no better than the decision tree technique.

## II A quick overview
Notwithstanding, the classic methods to tackle the fund selection and portfolio construction problems basically fall into following categories: the former looks at past 6mth or longer period returns, Sharpe ratios, alpha against certain index, etc; and the latter ranges from the classic tangent point on Markowitz/CAPM efficient frontier, to Eugene and Fama's 3 factors model, Barrar's risk model, Bridegewater's risk parity, and hedge fund's long short strategy, etc in more recent years.

The co-integration method, a step forward above the stationary time series theory, is simply linear regression on log price in practical words, preserving the time series structure over the observation period. DO NOT regress on returns! It relates to relative valuation and results to pair trading in practical investment. The key is to look at the evolving regression residual over time.

## III The models
First of all, we note the fundamental building block, the price return is approximated by log return as follow.
$$r = P_2/P_1 - 1 \approx \ln(P_2) - \ln(P_1)$$
The implication is that it translates division into subtraction, ie: return to log(price), and saves us the effort of dynamic rebalancing. Suppose we have two stocks, their price series are $\{P_i\}$, $\{Q_i\}$. Thus their return difference is simplified:
$$r_p - w \cdot r_q = (P_2/P_1 - 1) - w \cdot (Q_2/Q_1 - 1) = \ln(P_2) - w \cdot \ln(Q_2) - (\ln(P_1) - w \cdot \ln(Q_1)) = D_2 - w \cdot D_1$$
where w is the weight of $i^{th}$ stock.

We briefly show that it is easy to find counterexamples that regression on returns is inaccurate. Suppose we build an equal weight portfolio consists of 2 stocks A and

B. We let daily return series $r_t$ following normal iid: $\{r_t\} \sim N(0.2, 5) = norminv(rand(), 0.2, 5)/100$, prices $A_t = A_{t-1}*(1+r_t)$, $B_t = B_{t-1}*(1-r_t/2)$, let portfolio $C_t = (A_t+B_t)/2$, thus C's $r_{c,t} = (r_t - r_t/2)/2$. Now we plot $r_{c,t}$ against $C_t/C_{t-1} - 1$ in Excel chart, and find it deviates far from y=x in many cases (refresh to update rand() ). The main reason is due to lack of rebalancing, ie we shall plugin $w_{A,t} = A_t/(A_t+B_t)$ in $r_{c,t}$, however log(price) is insensitive to this.

**3.1 Fund selection**

Most investors have difficulty in choosing from a number of similar funds, such as sector funds, or theme funds with similar price pattern or top 10 holdings. We propose a method to tackle this problem. The basic idea is to pick a single fund which is equivalent (in co-integration terms) to the equal weight index of those candidate funds. The main argument is this: these funds are already carefully selected, with the poor performers banned from the pool in the first place. Thus we just need to pick the neutral one(s), ignore the outperforming or underperforming ones, because we believe those funds eventually will mean revert back to normal. A commonly used method Principle Component Analysis (PCA) [Fung and Hsieh 1997] follows the same idea.

*Algorithm 1  Fund selection*

**step1**: *Construct an equal weight portfolio C(t) for candidate funds, over past 1 year period t, daily frequency*

$$C_t = \frac{1}{N} \sum ln(P[i])_t \quad i=1,2,\ldots N$$

**step2**: *For each individual fund i, compute the regression residual a*

$$a[i](t) = C - \beta[i] * ln(P[i])$$

where $\beta[i]$ *is the linear regression coefficient of* $C \sim ln(P[i])$

**step3**: *Examine the chart of a[i](t), pick the one(s) exhibit white noise, i.e. mean=0 and variance is small*

**3.2 Fund replication**

Fund replication helps us to explain the source of fund return from its constituent holdings, which is very intuitive. On the other hand, the factor model is difficult to understand as the factors are not observable.

Secondly, it solves the liquidity problem. Most mutual funds have size limit. For example a 200 million size fund does not allow a 100m investor entering and exiting at frequent basis. The redemption fee is also high, 0.5-1%. Thus the investor has genuine interests to replicate the star funds by directly investing in equity market which has deep liquidity.

Specifically, every equity fund discloses its top 10 holdings in quarterly filings within 15 days after each quarter end by law. Naively replicating the fund by investing in its top 10 holdings at the disclosed weights will result to a disaster. We suggest performing a simple linear regression on a fund's top 10 holdings (from the latest quarterly filing for simplicity) to obtain new weights. This fund could be the one found in model 2. Some weights $\beta[i]$ will be negative, meaning that we shall have a long-short portfolio. Other replication methods include hedge fund replication from a number of macro factors such as stock indices, bond ETF and commodity ETF, and some newly developed techniques include stepwise selection methods, eg: Linear Clones Method [Hasanhodzic and Lo 2007], and Sequential Oscillating Selection Method [Byrd, et al. 2019] from larger stock pool.

Many long-short investors are puzzled with the situation that the total weights $\beta[i]$ sum up to nearly zero, suggesting the portfolio can be constructed at nearly zero cost. The real world is that in many exchanges, the money from short selling cannot be withdrawn until the positions are closed. Thus the actual money outlay is the sum of positive weights only, ie, sum($\beta > 0$). If the sum is less than 1, it means the remaining part (1-sum) will be deposited in Cash.

*Algorithm 2  Fund replication*

**step1**: *Regress the fund price series C(t) against top 10 holdings (latest quarter) of a fund over past 1 year period t, daily frequency:*

$$ln(C(t)) = \sum \beta[i] * ln(P[i]) + \varepsilon(t)$$

**step2**: *Examine the $\varepsilon(t)$ for stationary, ie: white noise whose mean=0 and variance small. (simply visual check the chart of $\varepsilon(t)$)*

**3.3 Optimal fund construction - the ultimate**

The replicated fund in model 2 still suffers from the overall market volatility, and will scare retail investors away in extreme market situation. Now we replace the left hand side of model 2 with the ultimate fund $e^{rt}$, a constant rate all weather portfolio, which is also very similar to hedge fund's NaV trend, especially the long-short strategy fund. Instead of picking stocks from a big universe such as the index constituents, we focus on the top 10 stocks of a sector fund or theme fund, because they are highly correlated. Thus the solution exists with high chance. Now the model 2 turns into model 3 as follow.

*Algorithm 3  optimal fund construction–the ultimate*

*step1: Regress the constant rate fund against the top 10 holding stocks (latest quarter) over past 1 year period t, daily frequency.. Initialize **r** =5% (eg: the investor desired rate at current market)*

$$r*t = \Sigma \beta[i] * ln(P[i]) + \varepsilon(t)$$

*step2: Examine the ε (t) for stationary, ie: white noise whose mean=0 and variance small. (simply visual check the chart of ε (t))*

*step3: Let S=sum(β >0), if S<1, it means (1-S) out of 1 dollar shall be deposited in Cash. While S>1 suggests borrowing on margin to invest.*

*Step4: Increase **r** to you desired level, and repeat the regression in Step 1. As a result, S will increase, meaning that we invest more in the long-short position and less in Cash.*

## IV The Data

We select a sample mutual funds for past 1 year period 2019Mar01-2020Apr01 in mainland China market to give a light taste of the general methodologies as shown above. Note the code for stocks and mutual funds in China is in 6 digit format. We program using the free online python environment (similar to quantopian.com) provided by the leading financial data vendor wind.com, the China version of Bloomberg.

For model 1, we showcase a number of interesting "carefully managed" fixed income funds which exhibit **y=e^rt** form of price trend in the long lasting bond bull market due to quantitative easing (QE) worldwide. The fund managers seem to use some earnings management technique, thus the funds' valuation is like the amortized cost accounting of high yield bond (6-8% p.a) whereas the risk free rate is 3-4%.

For model 2 and 3, we select a number of medical sector funds to illustrate our methodology.

## V Results and Discussion
### 5.1 Fund selection
We have 9 bond funds list below. The right panel is the price chart, plus a benchmark **y=e^{8%*t}** at the last. It is very difficult to pick a good one as their price trends look the same.

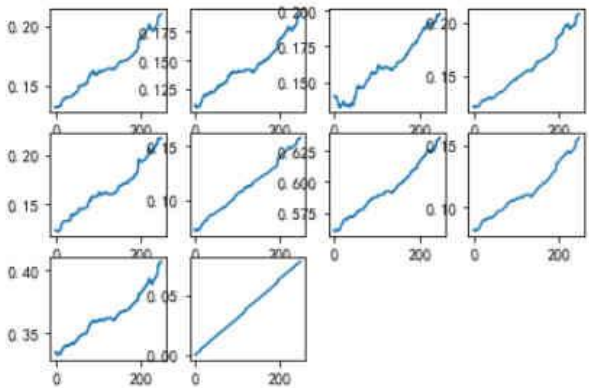

Fig1 Nine bond funds

Fig2 price series of the nine funds plus a benchmark

Below is the regression results of the equal weight portfolio against each funds (beta and $R^2$), while the Fig4 is plot of residue. We may pick No1, 3 or 5 as they exhibit white noise price pattern.

```
0  [1.0265]   r_2= 0.997
1  [1.0331]   r_2= 0.996
2  [1.1166]   r_2= 0.982
3  [0.865]    r_2= 0.993
4  [0.8428]   r_2= 0.995
5  [0.8803]   r_2= 0.989
6  [1.0588]   r_2= 0.998
7  [1.0705]   r_2= 0.997
8  [1.158]    r_2= 0.994
9  [0.9332]   r_2= 0.976
```

Fig3 regression result (beta)

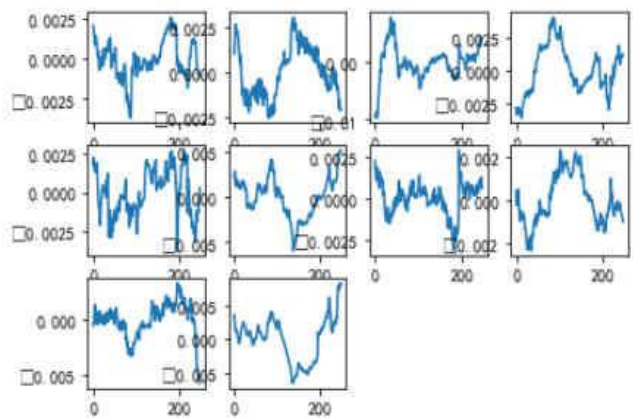

Fig4 residual plot

## 5.2 Fund replication

We select a stock fund ie: 000418 Invesco-GreatWall's Growth Stars fund. It has both uptrend and downtrend for the selected period. The replication results are shown below.

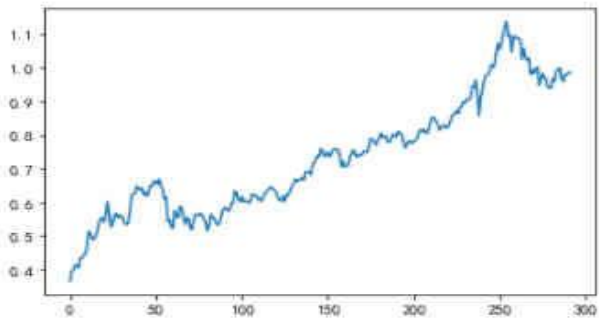
Fig5 top 10 holding (Mar31-2020)

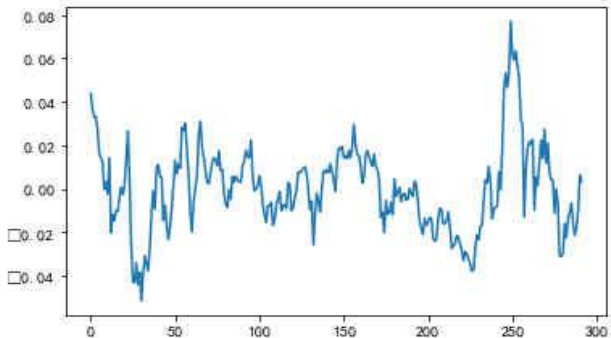
Fig6 price series of this fund

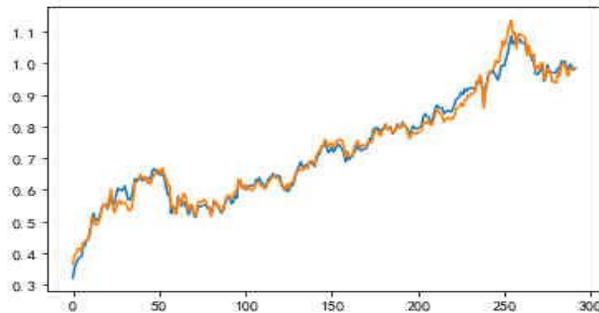
Fig7 plot of regression residual

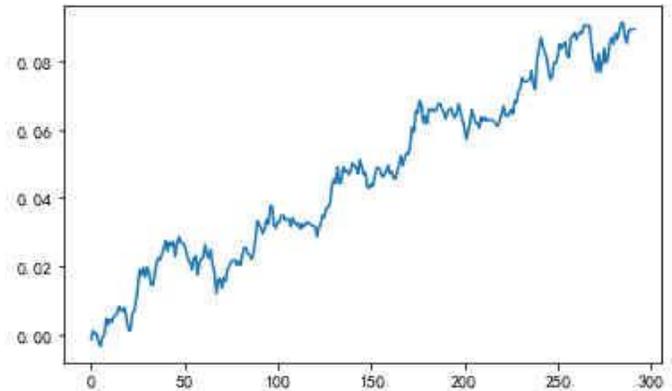
Fig8 plot of replicated fund and fund itself

beta= [ 0.2727  0.1039  0.1312  0.0854  0.1087 -0.0559  0.0519 -0.0988  0.1315  -0.0301]
$R^2$= 0.987

## 5.3 Optimal fund construction- the ultimate

Finally we regress the same fund's holding in section 5.2 against the ultimate fund: $y=e^{rt}$. Here we use r=8% for example.

beta*100= [-3.56, -2.31, -1.85, -1.71, -0.03, 0.48, 0.52, 3.71, 3.77, 8.4], $R^2$ = 0.961

sum(beta>0)= 16.8% means we have cash outlay of 16.8 out of 100 dollars in long position and 100-16.8=83.2 plus cash from short position in cash account. The portfolio grows at 8% p.a. on the total 100 dollars.

The price trend of the constructed portfolio is shown below.

Fig9 replicated fund, exhibiting $y=e^{rt}$ type of price trend

The model validity is based on the common assumption that the past can predict the near future, without regime change, which is essentially a momentum strategy. Refinement shall be developed when using in practical investment, eg: hold out 10% sample for out-of-sample validation, but it does not preclude us from generating alpha when others are scrutinizing boarder set of stocks/ETFs.